\documentclass[prl,twocolumn]{revtex4}
\usepackage{graphicx}

\begin{document}

\title{Continuous-Variable Quantum Information Distributor: Reversible Telecloning}
\author{Jing Zhang$^{\dagger }$, Changde Xie, Kunchi Peng}
\affiliation{State Key Laboratory of Quantum Optics and Quantum
Optics Devices, Institute of Opto-Electronics, Shanxi University,
Taiyuan 030006, P.R.China}

\begin{abstract}
We propose a scheme of continuous-variable reversible telecloning,
which broadcast the information of an unknown state without loss
from a sender to several spatially separated receivers exploiting
multipartite entanglement as quantum channels. In this scheme,
quantum information of an unknown state
is distributed into $M$ optimal clones and $M-1$ anticlones using $2M$%
-partite entanglement. For the perfect quantum information
distribution that is optimal cloning, $2M$-partite entanglement is
required to be a maximum two-party entanglement. Comparing with the
quantum telecloning proposed by Loock and Braunstein [Phys. Rev.
Lett. \textbf{87}, 247901 (2001)], this protocol produces the
anticlones (or time-reversed state) of the unknown quantum state,
thus, keep all information of an unknown state.
\end{abstract}

\maketitle

$Introduction.$ --- One of the main tasks in quantum information
processing and quantum computation is the distribution of quantum
information encoded in the states of quantum systems. The perfect
distribution requires the no loss of the quantum information of the
unknown state, that means this process is reversible and the unknown
state can be reconstructed in a quantum system again. It is now well
known that quantum information can not be exactly copied\cite{one}.
Although exact cloning is impossible, one can construct approximate
cloning machines. Buzek and Hillery proposed a universal quantum
cloning machine for an arbitrary quantum state where the copying
process is independent of the input states\cite{two}. In recent
years, quantum information and communication have been extended to
the domain of continuous variable (CV)\cite{four}, due to relative
simplicity and high efficiency in the generation, manipulation, and
detection of CV state. To date, the CV local cloning has been
studied intensively\cite{five,six,seven,eight,nine,ten}.

In parallel, quantum nonlocal cloning (telecloning) has also been
intensively studied, which is a combination of quantum cloning and
teleportation performed simultaneously. The aim of telecloning is to
broadcast information of an unknown state from a sender to several
spatially separated receivers exploiting multipartite entanglement
as quantum channels. For qubits, Bru$\beta $ et al. first proposed
$1\rightarrow 2$ telecloning, which use nonmaximum tripartite
entanglement (here it is named as irreversible teleclone
states)\cite {eleven}. In this case, the anticlones (phase-conjugate
clones, or time-reversed state) are lost, thus, quantum channel
don't require maximum entanglement. This kind of telecloning is
called irreversible telecloner and is regarded as imperfect nonlocal
distributor of quantum information. More generally, $1\rightarrow M$
irreversible teleclone states, which are $M+1$-partite entanglement,
are given in Ref.\cite{twelve}. Later, Murao et al. proposed a new
$1\rightarrow M+(M-1)$ telecloning scheme, in which quantum
information of an input qubit is distributed into $M$ optimal clones
and $M-1$ anticlones using $2M$-partite entanglement\cite{thirteen}.
This kind of telecloning is called reversible telecloner and is
regarded as perfect nonlocal distributor of quantum information. Due
to no loss of quantum information, $2M$-partite entanglement is
required to be a maximum two-partite entanglement. More generally,
qubit telecloning with $N$ identical inputs distributed among $M$
receivers has been studied\cite{forteen}. For continuous variables,
Loock and Braunstein proposed optimal $1\rightarrow M$ telecloning
of coherent states via a $M+1$-partite entangled state\cite
{sixteen}. It is emphasized in the protocol that optimal telecloning
can be achieved by exploiting nonmaximum bipartite entanglement
between the sender and all receivers. This result is not surprising
since the anticlones are not produced in this protocol and the
quantum information of the unknown state is lost in the process of
distribution. This scheme is regarded as the CV irreversible
telecloner and corresponds to the irreversible telecloner in the
domain of discrete variables\cite{eleven,twelve}. Furthermore, the
CV irreversible telecloning was studied in noise
environment\cite{seventeen}. Recently, irreversible telecloning of
optical coherent states was demonstrated
experimentally\cite{eighteen}. In this Letter, we propose a scheme
of CV reversible telecloning, which broadcast the information of an
unknown state without loss from a sender to several spatially
separated receivers exploiting multipartite entanglement as quantum
channels. In this process, quantum information of an unknown state
is distributed into $M$ optimal clones and $M-1$ anticlones using
$2M$-partite entanglement. For the reversible telecloning (the
perfect quantum information distribution) that is optimal cloning,
$2M$-partite entanglement used for quantum channel is required to be
a maximum two-party entanglement. Further, we generalize
$1\rightarrow M+(M-1)$ quantum telecloning to $N\rightarrow M+(M-N)$
case and also provide an explicit design of an asymmetric reversible
telecloning. Like the quantum teleportation, we give the lower and
upper bounds to achieve quantum telecloning when the imperfect
quantum entanglement is utilized.

$1\rightarrow 2+1$ $telecloning.$ --- A schematic setup for CV $1\rightarrow
2+1$ telecloning is depicted in Fig.1. The quantum states we consider in
this Letter can be described using the electromagnetic field annihilation
operator $\hat{a}=(\hat{X}+i\hat{Y})/2$, which is expressed in terms of the
amplitude $\hat{X}$ and phase $\hat{Y}$ quadrature with the canonical
commutation relation $[\hat{X},\hat{Y}]=2i$. Without a loss of generality,
the quadrature operators can be expressed in terms of a steady state and
fluctuating component as $\hat{A}=\langle \hat{A}\rangle +\delta \hat{A}$,
which have variances of $V_A=\langle \delta \hat{A}^2\rangle $ ($\hat{A}=%
\hat{X}$ or $\hat{Y})$. The heart of quantum telecloning is the
multipartite entanglement shared among the sender and receivers.
Without multipartite entanglement, it is only possible to perform
the corresponding two-step protocol: the sender produces clones and
anticlones locally, and then (bipartitely) teleports them to each
receiver. The two-step protocol would require $2M-1$ bipartite
entanglement for teleportation. Continuous-variable $1\rightarrow
2+1$ telecloning only needs one bipartite entanglement. The
bipartite entangled state of CV is two-mode Gaussian entangled state
(Einstein-Podolskyo-Rosen (EPR) entangled state), which can be
obtained directly by type-II parametric interaction\cite{ninteen} or
indirectly by mixing two independent squeezed beams on a
beam-splitter\cite{twenty}. The EPR entangled beams have the very
strong correlation property, such as both their
difference-amplitude quadrature variance $\langle \delta (\hat{X}%
_{a_{EPR1}}-\hat{X}_{a_{EPR2}})^2\rangle =2e^{-2r}$, and their
sum-phase quadrature variance $\langle \delta (\hat{Y}_{a_{EPR1}}+%
\hat{Y}_{a_{EPR2}})^2\rangle =2e^{-2r}$, are less than the quantum
noise limit, where $r$ is the squeezing factor. The EPR entangled
beams are divided into two beams at 50/50 beam splitters
respectively. The output modes $\hat{a}_{RTS1^{\prime }}$,
$\hat{a}_{RTS2^{\prime }}$, $\hat{a}_{RTS1}$ and $\hat{a}_{RTS2}$
are expressed as
\begin{eqnarray}
\hat{a}_{RTS1^{\prime }} &=&\frac{\sqrt{2}}2(\hat{a}_{EPR1}+\hat{v}_1),\quad
\hat{a}_{RTS2^{\prime }}=\frac{\sqrt{2}}2(\hat{a}_{EPR1}-\hat{v}_1), \\
\hat{a}_{RTS1} &=&\frac{\sqrt{2}}2(\hat{a}_{EPR2}+\hat{v}_2),\quad \hat{a}%
_{RTS2}=\frac{\sqrt{2}}2(\hat{a}_{EPR2}-\hat{v}_2),  \nonumber
\end{eqnarray}
where $\hat{v}_1$ and $\hat{v}_2$ refer to the annihilation
operators of the vacuum noises entering the beam splitters. This
output state is exactly Gaussian analog of $1\rightarrow 2+1$
reversible telecloning state of qubit when $r\rightarrow \infty $.
The $1\rightarrow 2+1$ telecloning state is partitioned into two
sets$\{{\hat{a}%
_{RTS1^{\prime }},\hat{a}_{RTS2^{\prime }}}\}$ and
$\{{\hat{a}_{RTS1},\hat{a}_{RTS2}}\}$. The parties in the same set
come from one of EPR entangled pair, so each party is in a thermal
state and shows excess noises, and there is no any quantum
entanglement between them. However, any two parties lying different
sets respectively have bipartitely entanglement. By
using the four-partite entangled modes, sender Alice can perform quantum $%
1\rightarrow 2+1$ telecloning of a coherent state input to three receivers
to produce two clones and one anticlone at their sites.

%
\begin{figure}
\centerline{
\includegraphics[width=3.3in]{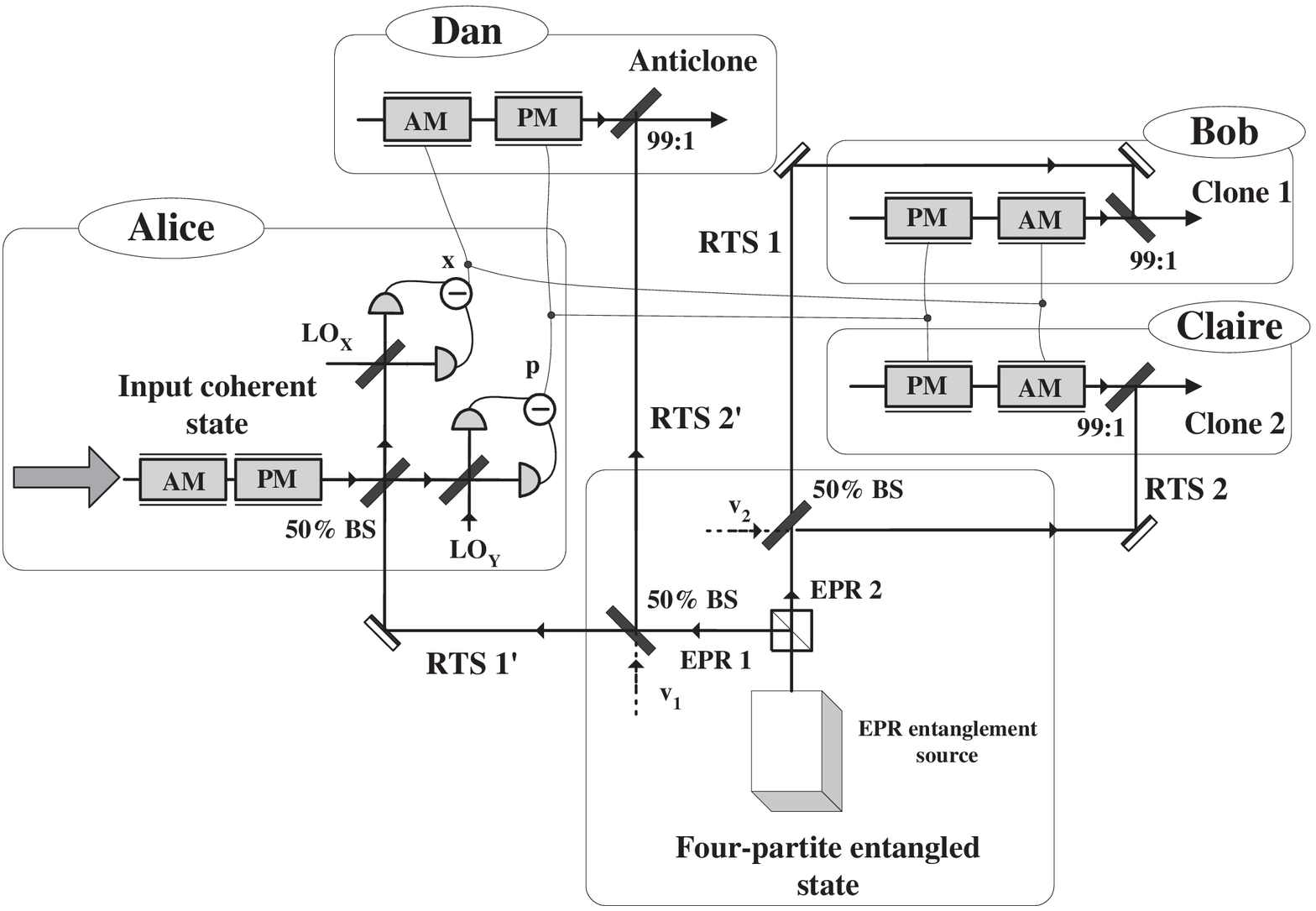}
} \vspace{0.1in}
\caption{ A schematic diagram of $1\rightarrow 2+1$ telecloning. BS:
Beam splitter, LO: Local oscillator, AM: Amplitude modulator, PM:
Phase modulator and AUX: Auxiliary beam. \label{Fig1} }
\end{figure}

For quantum $1\rightarrow 2+1$ telecloning, Alice first performs a
joint (Bell) measurement on her entangled mode
$\hat{a}_{RTS1^{\prime }}$ and an unknown input mode $\hat{a}_{in}$.
The Bell measurement consists of a 50/50 beam splitter and two
homodyne detectors as shown Fig.1. Alice's measurement results are
labeled as $x=(\hat{X}_{RTS1^{\prime }}-\hat{X}_{in})/\sqrt{2}$ and
$p=(\hat{Y}_{RTS1^{\prime }}+\hat{Y}_{in})/\sqrt{2}$. Receiving
these measurement results from Alice, Bob, Claire and Dan modulate
the amplitude and phase of an auxiliary beam (AUX) via two
independent modulators with the scaling factor $g_x^{B(C,D)}$, and
$g_p^{B(C,D)}$, respectively. The
modulated beams are combined with Bob, Claire and Dan's modes ($\hat{a}%
_{RTS1}$, $\hat{a}_{RTS2}$ and $\hat{a}_{RTS2^{\prime }}$) at 1/99
beam splitters. The output modes produced by the telecloning process
are expressed as
\begin{eqnarray}
\hat{a}_{out}^B &=&\hat{a}_{in}+\frac{\sqrt{2}}2(\hat{a}_{EPR2}-\hat{a}%
_{EPR1}^{\dagger })+\frac{\sqrt{2}}2(\hat{v}_2-\hat{v}_1^{\dagger }), \\
\hat{a}_{out}^C &=&\hat{a}_{in}+\frac{\sqrt{2}}2(\hat{a}_{EPR2}-\hat{a}%
_{EPR1}^{\dagger })-\frac{\sqrt{2}}2(\hat{v}_2+\hat{v}_1^{\dagger }),
\nonumber \\
\hat{a}_{out}^D &=&\hat{a}_{in}^{\dagger }-\sqrt{2}\hat{v}_1,  \nonumber
\end{eqnarray}
where we have taken $g_x^B=g_x^C=g_x^D=-\sqrt{2}$ and
$g_p^B=g_p^C=-g_p^D=\sqrt{2}$. From these equations, we can see that
Bob and Claire, whose entangled parties lie in different set with
Alice, get the cloned states. The cloned states have additional
noise terms to the input mode\cite{five}. This noise is minimized in
the case $r\rightarrow \infty $ corresponding to perfect EPR
entanglement. These are the optimal clones of coherent state input.
Dan is possessed of the entangled party lying in the same set with
Alice, so he achieves anticloned state, which has the complex
conjugate of the input state and the additional noise. This
additional noise is independent on the EPR entanglement. This always
is the optimal anticlone of coherent state input. In the case of
perfect EPR entanglement, the unknown input state is completely
unknown not only to Alice but to anyone in the process of
telecloning. Thus quantum information of the unknown state is
partitioned and distributed completely to Bob, Clair and Dan. The
optimal two clones and anticlone in Bob, Clair and Dan may is
reversed to the original unknown state in Alice by the same
reversible telecloning state. Bob, Clair and Dan perform the joint
(Bell) measurement respectively on their entangled modes and clones
(anticlone). Receiving these measurement results from Bob, Clair and
Dan, Alice displaces her entangled mode and can generate the
original unknown state. However, the unknown state can not be
reconstructed only with two optimal clones. It is worth noting that
the optimal two clones and anticlone in Bob, Clair and Dan
constitute a tripartite entangled state, which exactly corresponds
to the $1\rightarrow 2$ CV irreversible telecloning
state\cite{sixteen}.

In real experiment, a maximally EPR entangled state is not available because
of finite squeezing and inevitable losses. To assess the quality of
telecloning like teleportation, we apply the fidelity measure $\mathcal{F}%
\equiv \left\langle \psi ^{in}\left| \hat{\varrho}^{out}\right| \psi
^{in}\right\rangle $\cite{twenty}. In the case of unity gain, the
fidelity for the Gaussian states is simply given by $ \mathcal{F}=
2/{\sqrt{( 1+\langle \delta \hat{X}_{out}^2\rangle ) ( 1+\langle
\delta \hat{Y}_{out}^2\rangle ) }}.$ For the classical case of
$r=0$, i.e., the EPR beams were replaced by uncorrelated vacuum
inputs, the fidelity of Bob and Clair's outputs is found to be
$\mathcal{F}_{class}=1/2$\cite{twenty-one}, which corresponds to the
classical limit for coherent state cloning. The fidelity of Dan's
anticlone is $\mathcal{F}_{antic}=1/2$, which is independent on the
quantum entanglement. When they share quantum entanglement $r>0$,
the fidelity of the clones of Bob and Clair is
$\mathcal{F}_{clone}=2/(3+e^{-2r})$. It is
clearly shows that Bob and Clair get the clones with fidelity $\mathcal{F}%
_{clone}>1/2$, thus the quantum $1\rightarrow 2+1$ telecloning of coherent
states is deemed successful. Note that the optimal fidelity of $1\rightarrow
2+1$ coherent-state reversible telecloning is $2/3$ for the clones and $1/2$
for the anticlone, which requires the maximally EPR entangled state.

%
\begin{figure}
\centerline{
\includegraphics[width=3.3in]{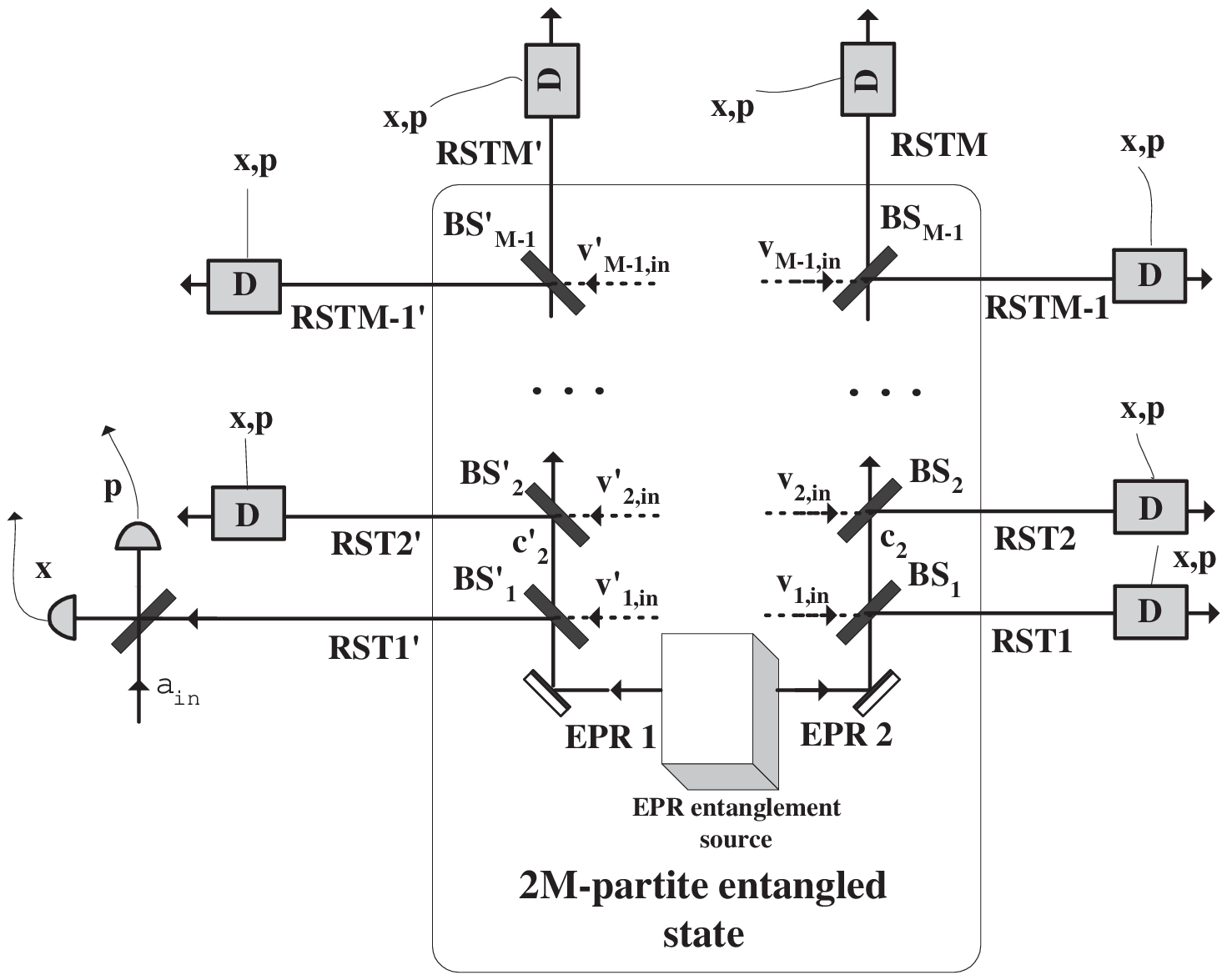}
} \vspace{0.1in}
\caption{ A schematic diagram of $1\rightarrow M+(M-1)$ telecloning.
\label{Fig2} }
\end{figure}

$1\rightarrow M+(M-1)$ $telecloning$. --- We now generalize
$1\rightarrow 2+1$ quantum telecloning to $1\rightarrow M+(M-1)$,
which produces $M$ clones and $M-1$ anticlones from a single input
state using $2M$-partite entanglement. We first generate the
$2M$-partite entanglement by a sequence of a EPR entangled beams and
$2(M-1)$ beam splitters with appropriately adjusted
transmittances and reflectances as illustrated in Fig.2. The modes $\hat{v}%
_{j,in}$ and $\hat{v}_{j,in}^{\prime }$ are in the vacuum state. The
EPR
entangled modes $\hat{a}_{EPR1}$ and $\hat{a}_{EPR2}$ are mixed with $\hat{v}%
_{1,in}$ and $\hat{v}_{1,in}^{\prime }$ at the beam splitters $BS_1^{\prime }
$ and $BS_1$, respectively. The mode $\hat{a}_{RTS1^{\prime }}$ $(\hat{a}%
_{RTS1})$ contains the EPR entangled mode $\hat{a}_{EPR1}$ ($\hat{a}_{EPR2}$%
) by a factor of $1/\sqrt{M}$. The output $\hat{c}_2^{\prime }$ ($\hat{c}_2$%
) is split at the $BS_2^{\prime }$ ($BS_2$) and so on, until we reach the
last beam splitter $BS_{M-1}^{\prime }$ ($BS_{M-1}$). The transformation
performed by the jth beam splitter can be written as
\begin{eqnarray}
\hat{a}_{RTSj^{(\prime )}} &=&\sqrt{\frac 1{M-j+1}}\hat{c}_j^{(\prime )}+%
\sqrt{\frac{M-j}{M-j+1}}\hat{v}_{j,in}^{(\prime )},
\label{tele-entanglement} \\
\hat{c}_{j+1}^{(\prime )} &=&\sqrt{\frac{M-j}{M-j+1}}\hat{c}_j^{(\prime )}-%
\sqrt{\frac 1{M-j+1}}\hat{v}_{j,in}^{(\prime )},  \nonumber
\end{eqnarray}
where $\hat{c}_1^{\prime }=\hat{a}_{EPR1}$, $\hat{c}_1=\hat{a}_{EPR2}$, and $%
\hat{a}_{RTSM^{(\prime )}}=\hat{c}_M^{(\prime )}$. It is clearly shows that
each $2M$-partite entangled mode $\hat{a}_{RTSj^{(\prime )}}$ (or $\hat{a}%
_{RTSj}$) contains $1/M$ portion of the the EPR entangled mode $\hat{a}%
_{EPR1}$ (or $\hat{a}_{EPR2}$) and $(M-1)/M$ portion of the vacuum
noise. The entanglement structure of $2M$-partite telecloning state
is also divided two sets $\{$ $\hat{a}%
_{RTS1^{\prime }},$ $\hat{a}_{RTS2^{\prime }},...,\hat{a}_{RTSM^{\prime }}\}$
and $\{$ $\hat{a}_{RTS1},$ $\hat{a}_{RTS2},...,\hat{a}_{RTSM}\}$. The
parties in the same set have no any quantum entanglement, however, any two
parties lying different sets respectively have bipartitely entanglement.

For quantum $1\rightarrow M+(M-1)$ telecloning, the sender chooses
any one of $2M$ modes of the telecloning state and performs a joint
measurement on his entangled mode and an unknown input mode
$\hat{a}_{in}$. Then the sender informs the other parties of its
measurement results $x$ and $p$. After receiving these measurement
results from sender, each party displaces its entangled mode by
modulating the amplitude and phase of an auxiliary beam, then
combining 1/99 beam splitter. The parties in the different set with
the sender produce the clones with $-g_x=g_p=\sqrt{2}$ and the
parties in the same set with the sender produce the anticlones with
$g_x=g_p=-\sqrt{2}$. The fidelity of $M$ clones and $M-1$ anticlones
is given by
\begin{eqnarray}
\mathcal{F}_{clone}^{1\rightarrow M+(M-1)} &=&\frac M{2M-1+e^{-2r}}, \\
\mathcal{F}_{antic}^{1\rightarrow M+(M-1)} &=&\frac 12.  \nonumber
\end{eqnarray}
The classical limit for $1\rightarrow M+(M-1)$ quantum telecloning is $%
\mathcal{F}_{class}=1/2$. The fidelity of the anticlones is $\mathcal{F}%
_{antic}=1/2$, which is independent on the quantum entanglement. When $r>0$,
the fidelity of the clones is larger than $1/2$, thus the quantum $%
1\rightarrow M+(M-1)$ telecloning of coherent states is successful. The $%
1\rightarrow M+(M-1)$ coherent-state telecloning become reversible and
optimal with the fidelity $M/(2M-1)$ for the clones and $1/2$ for the
anticlone when the EPR entangled state is perfect.

$N\rightarrow M+(M-N)$ $telecloning.$ --- We now address the most
complicated case, the $N\rightarrow M+(M-N)$ quantum telecloning,
which produces $M$ clones and $M-N$ anticlones from $N$ original
replicas of a coherent state using $2M$-partite entanglement. The
same multipartite entanglement Eq.\ref{tele-entanglement} is used
for the quantum channels. The $N$
replicas of a coherent state are stored the $N$ modes $\hat{a}_{in,1},...,%
\hat{a}_{in,N}$. In this scheme, we may consider to use a sender who holds
the $N$ input replicas and $N$ entangled modes in the same set of the $2M$%
-partite reversible telecloning state, or $N$ senders each of whom
holds one of $N$ input replicas and of the entangled modes in the
same set. Each input replica is performed the joint measurement with
a entangled mode. The
sender(s) generate $N$ amplitude- and phase-quadrature measurement results $%
(x_1,p_1),...,(x_N,p_N)$ and inform other parties. After receiving these
measurement results, each party first combines the measurement result $x_s=%
\frac{\sqrt{2}}N(x_1+...+x_N)$ and $p_s=\frac{\sqrt{2}}N(p_1+...+p_N)$, and
then displaces its entangled mode. The parties in the different set with the
sender produce $M$ clones with $-g_x=g_p=1$ and the parties in the same set
with the sender produce $M-N$ anticlones with $g_x=g_p=-1$. The fidelity of $%
M$ clones and $M-N$ anticlones is given by
\begin{eqnarray}
\mathcal{F}_{clone}^{N\rightarrow M+(M-N)} &=&\frac{NM}{NM+M-N+Ne^{-2r}}, \\
\mathcal{F}_{antic}^{N\rightarrow M+(M-N)} &=&\frac N{N+1}.  \nonumber
\end{eqnarray}
The classical limit for $N\rightarrow M+(M-N)$ quantum telecloning is $%
\mathcal{F}_{class}=N/(N+1)$. The fidelity of the anticlones is $\mathcal{F}%
_{antic}=N/(N+1)$, which is independent on the quantum entanglement. When $%
r>0$, the fidelity of the clones is larger than $N/(N+1)$, thus the quantum $%
N\rightarrow M+(M-N)$ telecloning of coherent states is successful. The $%
N\rightarrow M+(M-N)$ reversible telecloning requires the maximum EPR
entanglement, which is optimal cloner with the fidelity $MN/(MN+M-N)$ for
the clones and $N/(N+1)$ for the anticlone\cite{six}.

$Asymmetric$ $reversible$ $telecloning.$ --- Let us now demonstrate how to
make the reversible telecloning asymmetric. This is particularly interesting
in the context of quantum cryptography where it enables Eve to choose a
trade-off between the quality of her copy and the unavoidable noise that is
added to the copy sent to the receiver. Here we only concentrate on $%
1\rightarrow 2+1$ asymmetric telecloning. The scheme of $1\rightarrow 2+1$
asymmetric telecloning is similar to symmetric telecloning as in Fig.1, in
which only the vacuum noises $\hat{v}_1$ and $\hat{v}_2$ entering the beam
splitters are replaced by another EPR entangled beams $\hat{b}_{EPR1}$and $%
\hat{b}_{EPR2}$. Bob and Claire produce the clones and Dan produced the
anticlone, whose fidelity is written as
\begin{eqnarray}
\mathcal{F}_{clone}^B &=&\frac 2{2+e^{-2r}+e^{-2r_b}}, \\
\mathcal{F}_{clone}^C &=&\frac 2{2+e^{-2r}+e^{2r_b}},  \nonumber \\
\mathcal{F}_{antic}^D &=&\frac 2{2+e^{-2r_b}+e^{2r_b}},  \nonumber
\end{eqnarray}
where $r_b$ is the squeezing factor of the EPR entangled beams $\hat{b}%
_{EPR1}$and $\hat{b}_{EPR2}$. It clearly shows that $\mathcal{F}%
_{clone}^B>2/3>\mathcal{F}_{clone}^c$ and $\mathcal{F}_{antic}^D<1/2$ when  $%
r\rightarrow \infty $ corresponding to the reversible and optimal
telecloning. This means that the more Bob achieves the information of the
unknown state, the less Claire and Dan. The amount of information
distributed in the remote receivers is controlled by the squeezing factor $%
r_b$. This confirms that the device indeed realizes the optimal asymmetric
Gaussian telecloning of coherent states.

$Conclusion.$ --- We have introduced a scheme of CV reversible
telecloning, which broadcast the information of an unknown state
without loss from a sender to several spatially separated receivers
exploiting multipartite entanglement as quantum channels. In this
process, quantum information of an unknown state is distributed into
$M$ optimal clones and $M-1$ anticlones using $2M$-partite
entanglement. This new scheme of implementing quantum state
distribution nonlocally helps to deepen our understanding of the
properties of quantum communication systems enhanced by the
entanglement and its flexibility might have remarkable application
in quantum communication and computation.

$^{\dagger} $Corresponding author's email address: jzhang74@yahoo.com,
jzhang74@sxu.edu.cn

\textbf{ACKNOWLEDGEMENTS}

This research was supported in part by National Natural Science
Foundation of China (Approval No.60178012, No.60238010), and Program
for New Century Excellent Talents in University.

\end{document}